arXiv 2026

# Holistic law of aftershocks

A.V. Guglielmi

*Schmidt Institute of Physics of the Earth RAS, Moscow, Russia*
*guglielmi@mail.ru*

## Abstract

The paper is devoted to the phenomenological theory of aftershocks occurring in the source of a tectonic earthquake following the main shock. The theory was developed by the author jointly with A.D. Zavyalov and O.D. Zotov during the course of a long-term study of aftershocks. The theory is based on the concepts of source deactivation and the source's proper time. The holistic law governing the decay of aftershock activity over proper time follows from the theory. The damping decrement is equal to the source deactivation coefficient. The main focus of this paper is the analysis of the logical structure of the theory. The paper also contains a brief description of the experimental results obtained using the theory.

## Introduction

In the late 19th century, Omori proposed a theory regarding aftershocks that occur following the main earthquake shock. [*Omori*, 1894]. He introduced the concept of the instantaneous aftershock frequency $n(t)$ and experimentally substantiated the law of aftershock evolution

$$n(t) = k / t \ , \ t > 0 . \quad (1)$$

Here, $t$ is the time elapsed since the main shock, and $k$ is a fitting parameter. Formula (1) is known in geophysics as Omori's law.

In the early 20th century, Hirano noted that Omori's law is not holistic and proposed a more general formula

$$n(t) = k / t^{p} , \ \ p > 0 \quad (2)$$

to approximate experimental data [*Hirano*, 1924]. Utsu thoroughly investigated the applicability of formula (2) for the quantitative description of aftershocks [*Utsu*, 1961, 1962, 1971; *Utsu, Ogata, Matsu'ura*, 1995]. His work had a significant influence on the development of theoretical concepts regarding aftershocks. Formula (2) is commonly referred to as Utsu's law. Measurements from various regions of the globe indicate that, on average, $p = 1.1$, although the variation of the parameter $p$ from case to case is quite significant. Roughly speaking, the parameter ranges from $p = 0.7$ to $p = 1.5$. Research based on Utsu's methodology is documented in an extensive body of literature (see, for example, [*Ogata*, 1988; *Molchan, Dmitrieva*, 1992; *Wang*,1994; *Kisslinger*, 1996; *Ogata, Guo*, 1997; *Lombardi*, 2002; *Ogata, Zhuang*, 2006; *Holschneider et al.*, 2012; *Shcherbakov, Zhuang, Ogata*, 2018; *Rodrigo*, 2021; *Salinas-Martínez et al.*, 2023]).

The author proposed a holistic sigma-model of aftershocks based on concepts of deactivation and the proper time of the earthquake source [*Guglielmi*, 2016a; *Guglielmi*, 2016b]. A.D. Zavyalov, O.D. Zotov, and the author developed a method for studying the source based on the sigma model using aftershock observation data, and conducted an extensive experimental study of aftershocks. The results of years of collaborative work are presented in a series of culminating publications [*Guglielmi*, 2017; *Zotov et al.*, 2018; *Zavyalov, Guglielmi, Zotov*, 2020; *Zavyalov, Zotov*, 2021;

*Zavyalov et al.*, 2022; *Guglielmi et al.*, 2023, 2024, 2025; *Zotov, Guglielmi*, 2025a, 2025b].

The sigma model was conceived as an alternative to the traditional approach to the problem, which is based on the Omori and Utsu laws — major achievements in the physics of aftershocks from the 19th and 20th centuries. In a certain sense, the new approach is the direct opposite of the traditional one. The sigma model is designed to solve the inverse source problem, rather than the forward problem solutions to which are typically considered to be Omori's law or Utsu's law. The inverse problem consists of calculating the deactivation coefficient from experimental data on aftershocks. It is important to emphasize that, in constructing the sigma model, we adopted the concept of the instantaneous aftershock rate $n(t)$ from Omori and Utsu. However, unlike Omori and Utsu, we impose no restrictions on the form of the time dependence of the aftershock rate.

This paper focuses on the logical structure of the new theory. The essence of the sigma model is explained in an accessible manner. The author has endeavored to present the key ideas clearly and simply, aiming to provide a general overview without cluttering the picture with details.

Two circumstances drew our attention to the sigma model. First, the law of aftershock evolution takes on a simple and clear form when the source's proper time is used. A second factor is the high effectiveness of the methodology developed based on the model in identifying and investigating previously unknown dynamic properties of the source as it relaxes following the main shock.

## Sigma model

Let us base the theory of aftershocks on the following postulate:

*The aftershock rate $n(t)$ is a smooth, slowly decaying function of time.*

The condition of slowness implies that the relative change in the aftershock frequency over a time interval $T = 1/n$ is much less than unity. Mathematically, this is expressed by the strict inequality

$$n \gg \left|\frac{d}{dt}\ln n\right|. \qquad (3)$$

Let $\sigma$ be a small dimensionless parameter, and let $0 < \sigma \ll 1$. Then we can rewrite (3) in the form

$$\frac{dn}{dt} + \sigma n^2 = 0\,. \qquad (4)$$

Equation (4) follows quite logically from our postulate. We shall treat (4) as the equation governing aftershock evolution. We designate the parameter $\sigma$ as the earthquake source deactivation coefficient. The condition that the deactivation coefficient be small compared to unity is a necessary condition for the theory's applicability. Empirically, the condition $\sigma \ll 1$ is satisfied with a wide margin.

The question of the connection between the logical provability of the evolution equation and the physical content of the concept of deactivation cannot be resolved deductively. The issue was resolved experimentally by testing the hypothesis that the aftershock decay law is simple when earthquake timing is based on the source's proper time. We will return to this matter later and demonstrate that aftershocks decay exponentially with a decrement equal to the deactivation coefficient. The exponential decay of aftershocks lends physical definiteness to the source deactivation coefficient.

Following the traditional approach, one must specify the function $\sigma(t)$ based on certain considerations and then solve the Cauchy problem for the evolution equation (4). For example, for $\sigma$ = const, the solution takes the form of Omori's law (1). For $\sigma(t) = \text{const} \cdot t^{p-1}$, we obtain Utsu's law (2).

Our approach to the problem is diametrically opposed to the traditional approach. We shift the focus of attention from the aftershocks to the earthquake source and solve the inverse problem rather than the direct one. The essence of the inverse problem lies in calculating the source deactivation coefficient from experimental data on aftershock frequency. The correct solution to the inverse problem has the form

$$\sigma(t) = \frac{d}{dt}\langle g(t)\rangle . \qquad (5)$$

Here

$$g(t) = \frac{n(0) - n(t)}{n(0)n(t)} .$$

Angle brackets denote the optimal smoothing of the auxiliary function $g(t)$.

The sigma model admits interesting modifications and generalizations. For example, a change of variables leads to significant variants of the evolution equation. Substituting the dependent variable $n \to T = 1/n$ yields the simplest linear differential equation for aftershock evolution

$$\frac{d}{dt}T(t) = \sigma(t) \qquad (6)$$

instead of the nonlinear equation (4).

Let us introduce the concept of the proper time of the source

$$\tau(t) = \int_0^t \sigma(t')dt' . \qquad (7)$$

The use of proper time lends clarity and order to the picture of source relaxation. We see that the average waiting time for the next aftershock is a linear function of the proper time:

$$T(\tau) = T_0 + \tau . \qquad (8)$$

Let us recall that, in reality, aftershocks constitute a discrete sequence of seismic shocks. It is evident that, for constant $\sigma$, the time intervals between successive aftershocks form an increasing arithmetic progression. This statement is essentially a formulation of the classical Omori law (1) when aftershocks are viewed as a discrete sequence of earthquakes.

We view the source as a dynamic system. The parameter $\sigma$ describing the system must remain invariant under a change in aftershock chronometry. The invariance of σ upon transitioning from one method of timekeeping to another is not as obvious as it might seem at first glance. Let us provide some clarification on this matter.

Up to this point, we have measured proper time using imaginary clocks whose operating principle is described by formula (7). However, the synchronization of clocks displaying proper time with world time can be achieved in a different way. Let us perform a change of the independent variable: $t \to x(t)$. We shall call the inverse function $t(x)$ the synchronization function relating universal time $t$ and proper time $x$. (The symbol x comes from the word χρόνος.) The design of the "underground clock" in which the unit of time is the interval between two successive aftershocks dictates the choice of exponential synchronization:

$$t(x) = t_1 \exp \int_1^x \sigma(x') dx' . \qquad (9)$$

The aftershock equation takes the form

$$\frac{dn}{dx} + \sigma n = 0 . \qquad (10)$$

The deactivation coefficient is calculated using the synchronization function:

$$\sigma(x) = \frac{d}{dt} \langle \ln t(x) \rangle . \qquad (11)$$

It follows from the principle of sigma-invariance of the theory that $\sigma(t)=\sigma\left[x(t)\right]$. This equality can be used as a test when investigating the consistency of the sigma-theory.

## Successes of the Sigma Model

Our ten years of experience in the experimental study of aftershocks have demonstrated the utility of the sigma model. Let us outline the most interesting results.

**Relevance of the theory**. The sigma model is holistic and well-suited for problems in aftershock physics, as the measured deactivation coefficient is small compared to unity across a wide range of mainshock magnitudes, from M = 5 to M = 9. As the magnitude increases, the deactivation coefficient systematically decreases.

**The Omori epoch**. Omori's law holds true, but only during the initial stage of aftershock evolution; in other words, the law is not holistic. We have termed this first stage of evolution the "Omori epoch." The duration of the Omori epoch varies from case to case, ranging from a few days to several weeks. A trend is observed wherein the duration of the Omori epoch increases with the magnitude of the mainshock. For instance, following the Tohoku earthquake (M = 9.1), the deactivation coefficient remained constant at $\sigma$ = 0.0014 for four weeks. Subsequently, chaotic fluctuations in the deactivation coefficient set in, and Omori's law ceased to apply.

**Utsu's law**. During the first stage of aftershock evolution, Omori's law holds. During the second stage, non-monotonic variations in the deactivation coefficient are observed. Thus, observations do not support Utsu's law for any of the specific events analyzed, as the law predicts strictly monotonic behavior of the deactivation coefficient. (Recall that choice $\sigma(t)=\text{const}\cdot t^{p-1}$ corresponds to law (2).) Nevertheless, the experience of applying Utsu's law to describe aftershocks is exceptionally valuable. First, the law can be viewed as a compact fitting formula that provides a good approximation of observational data for a specific aftershock sequence. Second, experience with Utsu's law has shown that, averaged across the entire set of known measurements, $p = 1.1$. This value deviates only slightly from $p = 1$, the value at which Omori's law holds. Indeed, this difference can be considered negligible when accounting for the random and systematic errors inherent in measurements taken under the difficult conditions of field observation. These considerations prompted the development of the sigma-model and the formulation of the inverse problem regarding the source.

**Time relativism**. The proper time of an earthquake source elapses significantly more slowly than universal time. This property is evident from the fact that the deactivation coefficient is much smaller than unity. An important discovery is that proper time flows non-uniformly. The law governing aftershock evolution acquires particular simplicity and clarity when events are timed according to the source's proper time.

**Underground clocks**. A remarkable result was obtained by shifting from a continuous representation of the aftershock flow to a more realistic view of aftershocks as a discrete sequence of events. Using an earthquake catalog, one can assign an ordinal number (1,2,3, etc.) to each aftershock, as well as a corresponding Universal Time ($t_1$, $t_2$, $t_3$, etc.). If we plot these ordinals and times on a coordinate plane ($x$, $t$) and fit a smooth curve $x(t)$ to the points, the inverse function $t(x)$ can be defined as the function synchronizing Universal Time with the source's proper time. In other words, we have envisioned a giant underground clock that marks discrete moments of proper time through subterranean shocks.

**Exponential decay of aftershocks**. The simplicity and clarity of the aftershock law were achieved by using a "subsurface clock" to measure the source's proper time. It was found that aftershock activity decays exponentially with the passage of proper time. The decay decrement equals the source deactivation coefficient. The fundamental parameter of our theory has thus acquired physical definiteness.

**Triggers**. Let us generalize equation (4) by adding a term $f(t)$ to the right-hand side. The non-homogeneous differential equation simulates an additive trigger acting on the source. The impulse function $f(t)$ describes the cumulative effect of the seismic echo excited by the mainshock; this echo returns to the source approximately 3 hours after the mainshock and triggers aftershocks. The periodic function $f(t)$ describes the modulation of aftershocks by the Earth's spheroidal oscillations with a period of 54 minutes which are also excited by the mainshock.

**Foreshock activation**. Methods of analyzing experimental data, developed based on the sigma model, proved effective in the study of foreshocks. Foreshock activation was detected approximately $10^3$ hours prior to a magnitude $M \geq 5$ mainshocks.

**Convergence of foreshocks and divergence of aftershocks.** The convergence of foreshocks is evidenced by the fact that the average distance between foreshock epicenters and the mainshock epicenter decreases over proper time. The corresponding distance for aftershocks increases over proper time. This observed phenomenon indirectly indicates the existence of subsurface Umov energy flows.

## Discussion

The sigma model lends the theory of aftershocks the style characteristic of deductive science. We presented the evolution equation (4) as a consequence of a single axiom. Our elementary derivation (4) might seem like an impermissible trick, a kind of sophism. However, let us take into account that the law governing the evolution of aftershocks cannot be derived through strict logic. There is no avoiding conjecture and assumptions here. In other words, our phenomenological theory of aftershocks is not a strictly deductive theory. Along the way, we put forward heuristic hypotheses, that were by no means always obvious. Let us consider a simple example. Unlike Omori, we assumed that the parameter $\sigma$ is not a number determined empirically to best fit the experimental data to Omori's law. (In the classic works of Omori, Hirano, and Utsu, a constant value $k$ equal to $1/\sigma$ was used.) In contrast, we treat $\sigma$ as a function of time, $\sigma(t)$, and, crucially, interpret it as a key source parameter. The hypothesis of a time-varying source deactivation coefficient $\sigma(t)$ is arbitrary, but it is fully confirmed by experiment.

Thus, the fundamental postulate of the theory, cited above and enclosed in a box, should be regarded as a self-evident heuristic consideration that enabled the formulation of the dynamic equation (4). Clearly, one could have chosen other lines of reasoning, supported by heuristic hypotheses, that would lead us to law (4) or to an equivalent law of aftershock evolution. Let us present a few such possibilities here.

Omori's formula (1) plausibly describes source relaxation. Note that $dn/dt = \text{const} \cdot n^2$. A natural and entirely valid generalization is to replace the constant proportionality coefficient with the function $-\sigma(t)$. This yields equation (4).

An alternative approach is possible. We previously noted that the assertion that *the discrete sequence of intervals between successive aftershocks forms an increasing arithmetic progression* leads to the evolution equation in the form (4). The words in italics could be regarded as a postulate. Indeed, the graph of the arithmetic progression is the linear function $T(t) = T_0 + \sigma t$. A straightforward generalization immediately leads to equation (4).

A fascinating though logically more challenging path leads to the sigma model if one proceeds from the following postulate: *The frequency of aftershocks decays exponentially with the passage of proper time as measured by a subterranean clock*. We touched upon this issue in the previous section of the paper. We note here only that the proper-time intervals between two successive aftershocks form a stationary arithmetic progression, that is, a sequence of the form 1, 1, 1, … . As for the intervals of Universal Time synchronized with proper time, they form a geometric progression.

Thus, the plausibility of the sigma-model is evidenced not only by its success in uncovering new properties and patterns of aftershocks but also by the fact that the sigma-model can be constructed based on postulates that differ so greatly in form.

## Conclusion

We found that aftershock activity decays hyperbolically with the earthquake source's proper time $\tau$. When events are timed according to proper time $x$, aftershock activity decays exponentially with a decrement equal to the source deactivation coefficient $\sigma$.

An obvious drawback of our holistic theory is the use of the instantaneous aftershock rate $n$ to quantify the energy release accompanying the source relaxation process. We employed the concept of aftershock frequency introduced by the seismology pioneer Omori to maintain continuity when transitioning from the traditional approach to the aftershock problem to the new approach presented in the article as sigma-theory. Replacing frequency with mechanical energy output will require time and additional work.

***Acknowledgments.*** The Sigma model was developed and applied in practice through a long-standing collaboration with A.D. Zavyalov and O.D. Zotov. I express my deepest gratitude to them. I sincerely thank F.Z. Feygin and A.S. Potapov for their interest in the work and their support. This work was supported by the Ministry of Science and Higher Education of the Russian Federation within the framework of the state assignment of the O.Yu. Schmidt Institute of Physics of the Earth of the Russian Academy of Sciences.

## About the Author

Anatol Vladimirovich Guglielmi (born 1935). Education: Radiophysicist. Academic degree: Doctor of Physical and Mathematical Sciences. Academic rank: Professor of Geophysics. Position: Chief Researcher at the Schmidt Institute of Physics of the Earth, Russian Academy of Sciences (IPE RAS).